\documentclass[conference,onecolumn,12pt]{IEEEtran}
\ifCLASSOPTIONcompsoc
  \usepackage[nocompress]{cite}
\else
  \usepackage{cite}
\fi
\usepackage{amsmath}
\usepackage{algorithm}
\usepackage{url}
\usepackage{enumitem}
\usepackage{algorithmic}
\usepackage{graphicx}
\usepackage[export]{adjustbox}
\usepackage{todonotes}
\usepackage{hyperref}
\usepackage{verbatim}
\usepackage{footmisc}
\usepackage{multirow}
\usepackage{amssymb,amsmath}
\usepackage{amsthm}

\usepackage{flexisym}
\usepackage{subcaption}
\usepackage{textcomp}
\usepackage{xcolor}
\usepackage{makecell}

\graphicspath{{./images/}}

\newlist{problist}{itemize}{1}
\setlist[problist]{label=\textbf{P1}}

\begin{document}
%

\title{COVID-19 Risk Estimation using a Time-varying SIR-model}
%
%
%
%

\author{\IEEEauthorblockN{Mehrdad Kiamari, Gowri \\ Ramachandran, Quynh  Nguyen}
\IEEEauthorblockA{Viterbi School of Engineering,\\ University of Southern California,\\ Los Angeles, USA
    \\\{kiamari, gsramach, \\ quynhngu\}@usc.edu}
\vspace{-5mm}
\and
\IEEEauthorblockN{Eva Pereira,\\ Jeanne Holm}
\IEEEauthorblockA{Office of the Mayor, \\ City of Los Angeles, \\ Los Angeles, USA\\
\{eva.pereira, jeanne.holm\\\}@lacity.org}
\and
\IEEEauthorblockN{Bhaskar \\ Krishnamachari}
\IEEEauthorblockA{Viterbi School of Engineering\\, University of Southern California,\\ Los Angeles, USA
    \\\{bkrishna\}@usc.edu}
}


\IEEEtitleabstractindextext{%
\begin{abstract}
Policy-makers require data-driven tools to assess the spread of COVID-19 and inform the public of their risk of infection on an ongoing basis. We propose a rigorous hybrid model-and-data-driven approach to risk scoring based on a time-varying SIR epidemic model that ultimately yields a simplified color-coded risk level for each community. The risk score $\Gamma_t$ that we propose is proportional to the probability of someone currently healthy getting infected in the next 24 hours. We show how this risk score can be estimated using another useful metric of infection spread, $R_t$, the time-varying average reproduction number which indicates the average number of individuals an infected person would infect in turn. The proposed approach also allows for quantification of uncertainty in the estimates of $R_t$ and $\Gamma_t$ in the form of confidence intervals. Code and data from our effort have been open-sourced and are being applied to assess and communicate the risk of infection in the City and County of Los Angeles. 
 \end{abstract}

}


\maketitle
\IEEEpubidadjcol

\IEEEdisplaynontitleabstractindextext

%
\IEEEpeerreviewmaketitle

\section{Introduction}
  \label{sec:intro}
  
The ongoing COVID-19 epidemic has forced governments and public authorities to employ stringent measures~\cite{tobias2020evaluation,kupferschmidt2020lockdowns}, including closing business and implementing stay-at-home orders, to contain the spread. When making such decisions, policymakers require tools to understand in ``real-time" how the virus is spreading in the community, as well as tools to help communicate the level of risk to citizens so that they can be encouraged to take appropriate measures and take the public health directives seriously.

One metric that has been found to be useful for authorities to assess the level of containment over time is the effective reproduction number~\cite{liu2020reproductive}. The effective reproduction number, $R_t$, indicates on average how many currently susceptible persons can be infected by a currently infected individual. The epidemic grows if this measure is above one. It is desirable to keep this value as far below one as possible over time in order to contain and eventually, hopefully, eliminate the virus from the community. 

While $R_t$ is meaningful to understand the rate at which the epidemic is spreading and has been proposed previously (for example, see \url{https://rt.live/} ), what has been missing in the public discourse is a risk metric that is more suitable for communication to a wider public. One key requirement for such a metric is that it be something that a citizen could relate to on an individual basis. Another requirement is that it needs to be easy to communicate to a wide audience. We address both these requirements in this work and make the following contributions. 

First, we obtain the daily effective reproduction number $R_t$ of a time-varying SIR model as well as the corresponding confidence Interval. The confidence interval reflects uncertainty in both the parameter of the underlying model and uncertainty in the data itself. Further, we present the mathematical derivation of the distribution of $R_t$.

Second, we propose a novel risk score $\Gamma_t$ for a community that is proportional to the probability that an individual will get infected in the next 24 hours. We show that the risk score can be calculated given estimates of four quantities: a) an estimate of $I_{rep,new}(t)$, the most recently reported count of new confirmed infectious cases, b) an estimate of $R_t$ as discussed above, c) an estimate of $K$, the ratio of true infectious cases to the number of confirmed cases, and d) an estimate of $S(t)$, the current number of susceptible individuals in the community. To make the score more meaningful, we normalize the probability of infection by multiplying it by 10,000. Then, a risk score of $x$ is an indication that there is, on average, a chance of $x$ in 10,000 of an individual in the community becoming infected in the next 24 hours. 

Third, we propose to convert the numerical risk score, which has an intuitive meaning as indicated above, to a color-coded risk level based on suitably chosen thresholds. We propose the use of four color-levels to indicate the corresponding risk level from low to high: green, yellow, orange, and red. 

\begin{figure*}[t]
\centering
\includegraphics[scale=0.5]{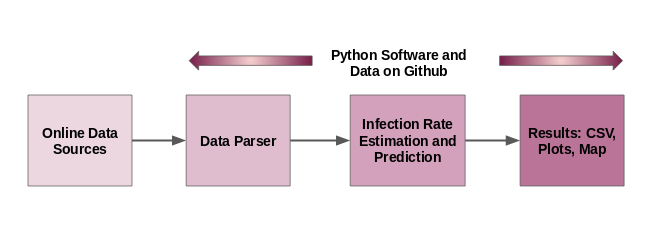}
\caption{Overview of Our System.}
\label{fig:system}
\end{figure*}

Fourth, we have implemented software to estimate the risk level for any community and released it as open-source. The code requires only time-series data on confirmed new cases, the population of the community, and an estimate for the ratio of true to confirmed (detected) COVID-19 positive cases. This software is being used at USC to process the daily data of communities within Los Angeles County to estimate and generate maps of risk levels by community. The block diagram in figure~\ref{fig:system} illustrates key elements of our system design. Our data parser is able to get the raw data from online data sources, clean them up and store them in machine-friendly (csv and json) formats. Our code for infection risk calculation uses this data in conjunction with a time-varying SIR-based Bayesian mathematical model to obtain risk estimates and prediction for different communities. The results are provided in CSV format and can be used to generate a heatmap-type visualization as well. 


The risk scoring model we describe in this work is now being used by the City of Los Angeles, which in turn is working with the County of Los Angeles and other partners to develop a publicly accessible tool that can be used by individuals and communities to grow awareness and mitigate risk of infection.  We believe that our risk estimation approach will be similarly of value to other communities around the world.

\section{Related Work}

As noted above, the calculation of the risk score requires an estimate of $R_t$. We show how this can be estimated using a time-varying SIR model, a generalization of the well-known SIR compartmental model~\cite{brauer2005kermack, sir} which consists of three states, namely the susceptible state,  the infected state,  and the recovered state. While traditionally this model is assumed to have a interaction rate / infection rate parameter that is constant, one recent work has used a time-varying SIR model to recover the time-varying effective reproduction number~\cite{timeseries}. Going beyond that work, we also show how to derive a confidence interval for $R_t$ in this work. Further, the authors of~\cite{timeseries} make strong assumptions on the number of susceptible individuals by approximating it as a constant factor of the entire population. This assumption may not be accurate when the number of infected individuals are high compared to the total population of a community; we therefore take a more general approach. 

Another recent work by Systrom~\cite{systrom_2020} has presented a Bayesian prediction approach to obtain confidence intervals for $R_t$. However, Systrom's work builds on~\cite{original_paper_Baysian_est}, where the definition of infection rate $R_t$ is not based on a time-varying contact rate of the SIR model. Instead, their approach estimates infection rate probabilistically based on the number of new cases alone. 

We are not aware of prior work that has proposed defining risk for COVID-19 or other epidemics in terms of an individual's probability of infection, which we argue is more meaningful for communicating risk to the public.

\section{Methodology}


Compartmental mathematical models for epidemic spreads including the well-known SIR model have been used since the work of Kermack and McKendrick in 1927~\cite{brauer2005kermack}. In the SIR model, each member of a given population is in one of three states at any time: susceptible, infectious, recovered. Any individual that is susceptible could become infected with some probability when they come into contact with an infected individual. Any individual that is infectious eventually recovers (in the context of COVID-19 when applying the SIR model, note that the category of recovered individuals will also include removed individuals due to deaths, which could be modeled as a constant fraction of all individuals in this category). In the classical SIR model, the number of susceptible individuals that become infected depends on the rate at which infected and susceptible individuals encounter each other and this rate is assumed to be constant. A well-known parameter in the classical SIR model is called R0, the effective reproductive number, which measures the average number of infections caused by infectious individuals at the beginning of the epidemic. 

\subsection{Time-Varying SIR model and $R_t$}
In our work, we have extended the SIR model to a time-varying model, in which the rate of encounters and infection probability between individuals in the population is assumed to be time-varying. This better reflects the reality of our present epidemic where interventions such as stay-at-home have been put in place and relaxed and various times and compliance with recommendations such as wearing masks and maintaining physical density has also been time-varying.  Based on this model, we are able to define and derive a new approach to calculating a time-varying version of the effective reproductive number, which we refer to as $R_t$. 

A particularly innovative aspect of our model is that it is a Bayesian model that allows the incorporation of various sources of uncertainty in the model, including uncertainty in the actual numbers of infected individuals (due to not every infected individual having been tested, as studies [2] have shown), uncertainty in recovery times, and uncertainty in the choice of parameters for de-noising the empirical data. This allows us to generate not only an estimate of $R_t$, but also quantify confidence in the estimate from a rigorous statistical perspective.

In this section, we elaborate upon the SIR model in detail. 
The SIR model is one of the simplest and the most well-known epidemic model~\cite{brauer2005kermack, sir} where each person belongs to one of the following three states: the susceptible state,
the infected state, and the recovered state. Regarding the susceptible state, individuals have not had the virus yet. However, they may get infected in case of being exposed to an infected individual. 
As far as the infected state is concerned, a susceptible person has the virus after being exposed to infected individuals. Finally, a person enters the recovered state in case of either the individual gets healed or dies. One important point about this model is that a recovered person will not be a susceptible one anymore.  

The SIR model follows the following differential equations:
\small
\begin{equation}\label{sir_diff}
\begin{aligned} 
\frac{dS(t)}{dt} &= -\beta \frac{S(t)I(t)}{N}\\
\frac{dI(t)}{dt} &= \beta \frac{S(t)I(t)}{N} - \sigma I(t)\\
\frac{dR(t)}{dt} &= \sigma I(t)
\end{aligned}    
\end{equation}
\normalsize
where $S(t)$, $I(t)$, and $R(t)$ respectively represent the number of susceptible, infected, and recovered people in a population size of $N$ at time $t$. Regarding the parameter $\sigma$, it is the recovery rate after being infected and is equal to $\frac{1}{D_I}$ where $D_I$ represents the average infectious days. Parameter $\beta$ is known as the effective contact rate, i.e. the average number of contacts an individual have with others is $\beta$. 

In analyzing whether any pandemic is contained, it is very crucial to obtain parameter $\beta$. We next show that how we can derive $\beta$ from the aforementioned differential equations.

\subsubsection{\bf Obtaining $\beta_t$ and $R_t$ for the SIR Model}
In the SIR model, we can express the number of susceptible individuals in terms of population size and the number of infected persons as $S(t) \approx N - I(t)$.
By replacing $S(t)$ with $N- I(t)$ in the second differential equation of (\ref{sir_diff}), we would have
\small
\begin{equation}\label{sir_diff_dI}
\begin{aligned} 
\frac{dI(t)}{dt} &= \beta \frac{\Big(N- I(t)\Big)I(t)}{N} - \sigma I(t).
\end{aligned}    
\end{equation}
\normalsize
We can rewrite (\ref{sir_diff_dI}) as follows:
\small
\begin{equation}\label{sir_diff_dI_integrable}
\begin{aligned} 
\frac{dI(t)}{(\beta-\sigma)I(t)-\frac{\beta}{N}I^2(t)} &= dt.
\end{aligned}    
\end{equation}
\normalsize
By taking definite integral from time $t_1$ to $t_2$ and assuming $\beta$ to be constant in this time interval, we would have
\small
\begin{equation}\label{sir_diff_dI_cal_integ}
\begin{aligned} 
\int_{t_1}^{t_2}{\frac{dI(t)}{(\beta-\sigma)I(t)-\frac{\beta}{N}I^2(t)}} &= \int_{t_1}^{t_2}{dt}
\end{aligned}    
\end{equation}
\normalsize
which leads to 
\small
\begin{equation}\label{sir_diff_dI_cal_integ_res}
\begin{aligned} 
\frac{1}{\beta-\sigma} \Big(\log \frac{I(t_2)}{\beta-\sigma-\frac{\beta}{N}I(t_2)}
-\log \frac{I(t_1)}{\beta-\sigma-\frac{\beta}{N}I(t_1)}
\Big) &= t_2-t_1
\end{aligned}    
\end{equation}
\normalsize

One can easily check (\ref{sir_diff_dI_cal_integ_res}) has a unique solution for $\beta$ due to the fact that term $\frac{1}{\beta-\sigma}$ and log term have monotonic behaviors.

An epidemic happens in case of increase in the number of infected individuals, i.e. $\frac{dI(t)}{dt}>0$, or consequently
\small
\begin{equation}\label{sir_diff_dI_positive}
\begin{aligned} 
\beta \frac{\Big(N- I(t)\Big)I(t)}{N} - \sigma I(t)& >0.
\end{aligned}    
\end{equation}
\normalsize
In the early stage of an epidemic, almost everyone are susceptible except very few cases. Therefore, $N- I(t)\approx N$ and as a result, condition (\ref{sir_diff_dI_positive}) would turn into $\frac{\beta}{\sigma} >1$.

 The variable $R\triangleq \frac{\beta}{\sigma}$ is defined as the \emph{effective reproduction number}. It is a useful metric to determine epidemic growth. In case of having $R > 1$, the epidemic is growing exponentially while $R<1$ indicates the epidemic is contained and will decline and die out eventually. 

For discrete-time cases such as daily reporting on number of infected cases, the time-variant effective contact rate $\beta_t$, which represents the contact rate for time slot $t$ can be derived by solving the following equation: 
\small
\begin{equation}\label{sir_diff_dI_cal_integ_res_disc}
\begin{aligned} 
\frac{1}{\beta_t-\sigma} \Big(\log \frac{I(t+1)}{\beta_t-\sigma-\frac{\beta_t}{N}I(t+1)}
-\log \frac{I(t)}{\beta_t-\sigma-\frac{\beta_t}{N}I(t)}
\Big) &= 1~\forall t.
\end{aligned}    
\end{equation}
\normalsize
Therefore, the time-variant effective reproduction number would be defined as $R_t \triangleq \frac{\beta_t}{\sigma}$.
Since it is difficult to write a closed form solution for $\beta_t$ in (\ref{sir_diff_dI_cal_integ_res_disc}), we take a simpler approximation to $\beta_t$ by considering the following which is based on (\ref{sir_diff_dI}) 
\small
\begin{equation}\label{sir_diff_dI_CI}
\begin{aligned} 
\beta_t &\approx \frac{\sigma I(t)+\Big(I(t+1)-I(t)\Big)}{\big(1-\frac{I(t)}{N}\big)I(t)}.
\end{aligned}    
\end{equation}
\normalsize

Then, we estimate $R_t$ as $\frac{\beta_t}{\sigma}$.

\subsubsection{\bf Obtaining the Confidence Interval for $R_t$}
Since there is uncertainty about parameter $D_I$ (or equivalently $\sigma$) and the number of infected cases $I(t)$, we now provide the derivation of confidence interval for parameter $R_t$. 
Regarding modeling the ambiguity in the number of the infected cases, we present the uncertainty about the actual number of infected cases as a factor of reported ones, i.e. $I_{rep}(t) \triangleq \frac{1}{K} I(t)$, and $K$ is a constant greater than 1. 
The main intuition behind this factor is due to taking into account the following two phenomena, namely lack of sufficient number of tests (specially in the beginning of the pandemic) and asymptomatic cases (mild infections which might not even be noticed).

To derive the confidence interval, we need to first find the marginal distribution of $R_t$.
By considering $f_{D}(d)$ and $f_{K}(k)$ as the probability distribution function (pdf) for parameters $D_I$ and $K$, respectively, the joint pdf of these parameters would be 
\small
\begin{equation}\label{joint_sigma_k}
\begin{aligned} 
f_{D,K}(d,k) & = f_{D}(d)f_{K}(k)
\end{aligned}    
\end{equation}
\normalsize
due to the independence of $D_I$ and $K$. 
We can derive the probability distribution function of $R_t$ by performing the following transformation on parameters $D_I$ and $K$ and introducing auxiliary variable $Z$:
\small
\begin{equation}\label{transformation}
\begin{aligned} 
Z & \triangleq K~~~~,~~~~
R_t & = \frac{1}{1-\frac{KI_{rep}(t)}{N}}\Big(1+D_I\frac{I_{rep}(t+1)-I_{rep}(t)}{ I_{rep}(t)} \Big).
\end{aligned}    
\end{equation}
\normalsize
Since the transformation of $(Z,R_t)$ to $(D_I,K)$ is one-to-one, we have
\small
\begin{equation}\label{transformation_inv}
\begin{aligned} 
K & = Z~~~~,~~~~
D_I & = \frac{R_t(1-Za_t)-1}{b_t},
\end{aligned}    
\end{equation}
\normalsize
where $a_t \triangleq \frac{I_{rep}(t)}{N}$ and $b_t \triangleq \frac{I_{rep}(t+1)-I_{rep}(t)}{I_{rep}(t)}$, the joint pdf of $Z$ and $R_t$ would be $f_{Z,R_t}(z,r)  = |J|f_{D,K}(d,k)$ 
with
\small
\begin{equation}\label{jaccobian}
\begin{aligned} 
J &\triangleq \begin{bmatrix}
\frac{\partial d}{\partial z} & \frac{\partial d}{\partial r} \\
\frac{\partial k}{\partial z} & \frac{\partial k}{\partial r} 
\end{bmatrix}.
\end{aligned}    
\end{equation}
\normalsize
By substituting the corresponding values of parameters and the Jacobin, we have: 
\small
\begin{equation}\label{joint_z_R_simplified}
\begin{aligned} 
f_{Z,R_t}(z,r) & = |\frac{1-za_t}{b_t}| f_{D}(\frac{r(1-za_t)-1}{b_t})f_{K}(z).
\end{aligned}    
\end{equation}
\normalsize
The marginal pdf of $R_t$ can be obtained by taking integral of (\ref{joint_z_R_simplified}) over parameter $z$, i.e.
\small
\begin{equation}\label{dist_R_appendix}
\begin{aligned} 
f_{R_t}(r) & = \int f_{Z,R_t}(z,r)dz
= \int |\frac{1-za_t}{b_t}| f_{D}(\frac{r(1-za_t)-1}{b_t})f_{K}(z)dz.
\end{aligned}    
\end{equation}
\normalsize

{\bf Remark 1}: One reasonable assumption regarding the pdf of parameters $D_I$ and $K$ is that both of them have Gaussian distributions. By considering $D_I \sim \mathcal{N}(\mu_D,\sigma_D^2)$ and $K\sim\mathcal{N}(\mu_K,\sigma_K^2)$, the pdf of $R_t$ can be simplified as 
\begin{equation}\label{dist_R_both_normal}
\begin{aligned} 
f_{R_t}(r) 
& = \int_{-\infty}^{\frac{1}{a_t}} (\beta_0+\beta_1z)C\sqrt{2\pi\sigma_c^2}\phi_{\mu_c,\sigma_c^2}(z)dz +
\int_{\frac{1}{a_t}}^{\infty} (-\beta_0-\beta_1z)C\sqrt{2\pi\sigma_c^2}\phi_{\mu_c,\sigma_c^2}(z)dz,
\end{aligned}    
\end{equation}
where $\phi_{\mu_c,\sigma_c^2}(.)$ indicates the pdf of a normal distribution with mean $\mu_c$ and variance $\sigma_c^2$ while
\begin{equation}\label{def_for_both_normal}
\begin{aligned} 
\beta_0  \triangleq \frac{1}{b_t}~~&,~~\beta_1 \triangleq \frac{-a_t}{b_t},\\
\alpha_0 \triangleq \frac{(\frac{r-1}{b_t}-\mu_D)^2}{2\sigma_D^2}+\frac{\mu_K^2}{2\sigma_K^2}
~~,~~\alpha_0 \triangleq &\frac{(-\frac{ra_t}{b_t})(\frac{r-1}{b_t}-\mu_D)}{\sigma_D^2}-\frac{\mu_K}{\sigma_K^2}
~~,~~\alpha_2 \triangleq \frac{(\frac{ra_t}{b_t})^2}{2\sigma_D^2}+\frac{1}{2\sigma_K^2},\\
\mu_c \triangleq \frac{-\alpha_1}{2\alpha_2}
~~,~~\sigma_c^2 \triangleq &\frac{1}{2\alpha_2}~~,~~C \triangleq \frac{e^{-(\alpha_0-\frac{\alpha_1}{4\alpha_2})}}{2\pi \sigma_D \sigma_K}.
\end{aligned}    
\end{equation}
By taking integral through using change of parameters, (\ref{dist_R_both_normal}) can be rewritten as follows
\begin{equation}\label{simplified_for_both_normal}
\begin{aligned}
f_{R_t}(r) & = -2C\beta_1 \sigma_c^2 e^{-\frac{(\frac{1}{a_t}-\mu_c)^2}{2\sigma_c^2}} +
C\sqrt{2\pi\sigma_c^2}(\beta_1\mu_c+\beta_0)\Phi_{\mu_c,\sigma_c^2}(\frac{1}{a_t})+C\sqrt{2\pi\sigma_c^2}(-\beta_1\mu_c-\beta_0)(1-\Phi_{\mu_c,\sigma_c^2}(\frac{1}{a_t}))
\end{aligned}    
\end{equation}
where $\Phi_{\mu_c,\sigma_c^2}(.)$ represents the cumulative distribution function (cdf) of a normal distribution with mean $\mu_c$ and variance $\sigma_c^2$. 

The confidence interval would belong to $({\bar R}_{t}-\delta,{\bar R}_{t}+\delta)$ where ${\bar R}_{t} \triangleq \mathbb{E}[{R}_t]=\int rf_{R_t}(r)dr$ and $\delta$ can be derived by satisfying $\mathbb{P}(|R_t-{\bar R}_t|\leq \delta)= \int_{{\bar R}_{t}-\delta}^{{\bar R}_{t}+\delta}{f_{R_t}(x)dx} = 1- \epsilon$
for some small $\epsilon>0$.

\subsubsection{\bf Estimating the Risk Score}


We propose a novel risk score metric for a given community that is proportional to the probability of someone in that community becoming infected in the next time period (typically, 24 hours). The risk score can be derived as the average number of people in that community that are likely to get infected in the next 24 hours by the currently infectious people divided by the current number of susceptible individuals. We further normalize this probability by multiplying by 10,000,  so that a score of 1 implies a 1 in 10,000 chance of getting infected, a score of 2 implies a 2 in 10,000 chance of getting infected, and so on. Mathematically, the risk score  is defined as follows: 


\begin{equation}
\Gamma_t = \dfrac {I(t) \cdot  R_t}{D_I \cdot S(t) }  \cdot 10000 \approx \dfrac {K \cdot I_{rep,new}(t) \cdot R_t}{N} \cdot 10000, 
\end{equation}
where $I_{rep,new}(t)$ indicates the most recently reported count of new confirmed infectious cases, $K$ refers to the ratio of true cases to reported cases, $R_t$ is the time-varying reproduction number, and $N$ is the total population size of the community. The approximation follows from the fact that $I_{rep,new}(t)$ is approximately equal to $\frac{I(t)}{D_I \cdot K}$ and $S(t)$ the number of susceptible people in the community is approximately equal to $N$ in the early stages of the epidemic. Confidence intervals for the risk score $\Gamma_t$ could be obtained numerically using a similar process as described for $R_t$ accounting also for uncertainty in $K$. Note that since $K$ may not be known for a given community, it may be helpful to use the following normalized form of the risk score: $\frac{\Gamma_t}{K}$, which is still proportional to the probability of infection for an individual.




\subsubsection{\bf Color-coded Risk Levels}
To further simplify the presentation of the risk score to a wider audience, we propose to classify the risk levels into four color-coded levels: (Green, Yellow, Orange, Red). The risk level is determined by evaluating the normalized risk score ($\frac{\Gamma}{K}$)  with respect to three pre-specified threshold levels $\theta_1, \theta_2, \theta_3$, such that when 
$\frac{\Gamma}{K} < \theta_1$ the risk level is green, when $\theta_1 \le \frac{\Gamma}{K} < \theta_2$ the risk level is yellow, when $\theta_2 \le \frac{\Gamma}{K} < \theta_3$ the risk level is orange, and when $\frac{\Gamma}{K} \ge \theta_3$ the risk level is red. 


\begin{figure*}[tbh]
\centering
\includegraphics[scale=0.4]{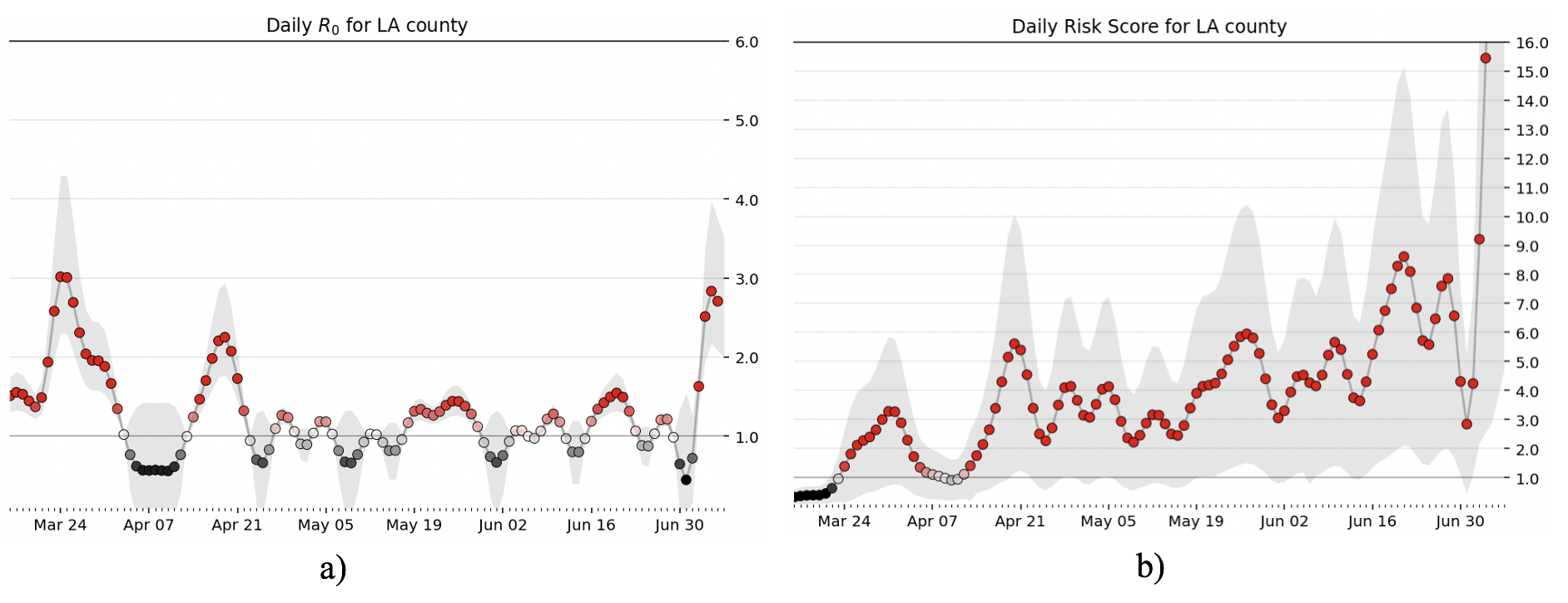}
\caption{ Plots a) and b) respectively represent the estimated effective reproduction number $R_t$ and the risk score $\Gamma_t$ over time for the entire county of LA considering $\mathbb{E}[D_I]=7.5$, $Var[D_I]=9$,  $\mathbb{E}[K]=3$, and $Var[K]=0.44$. The gray area represents the 95\% confidence interval in the estimates.}
\label{fig:entire}
\end{figure*}

\begin{figure*}[tbh]
\centering
\includegraphics[scale=0.35]{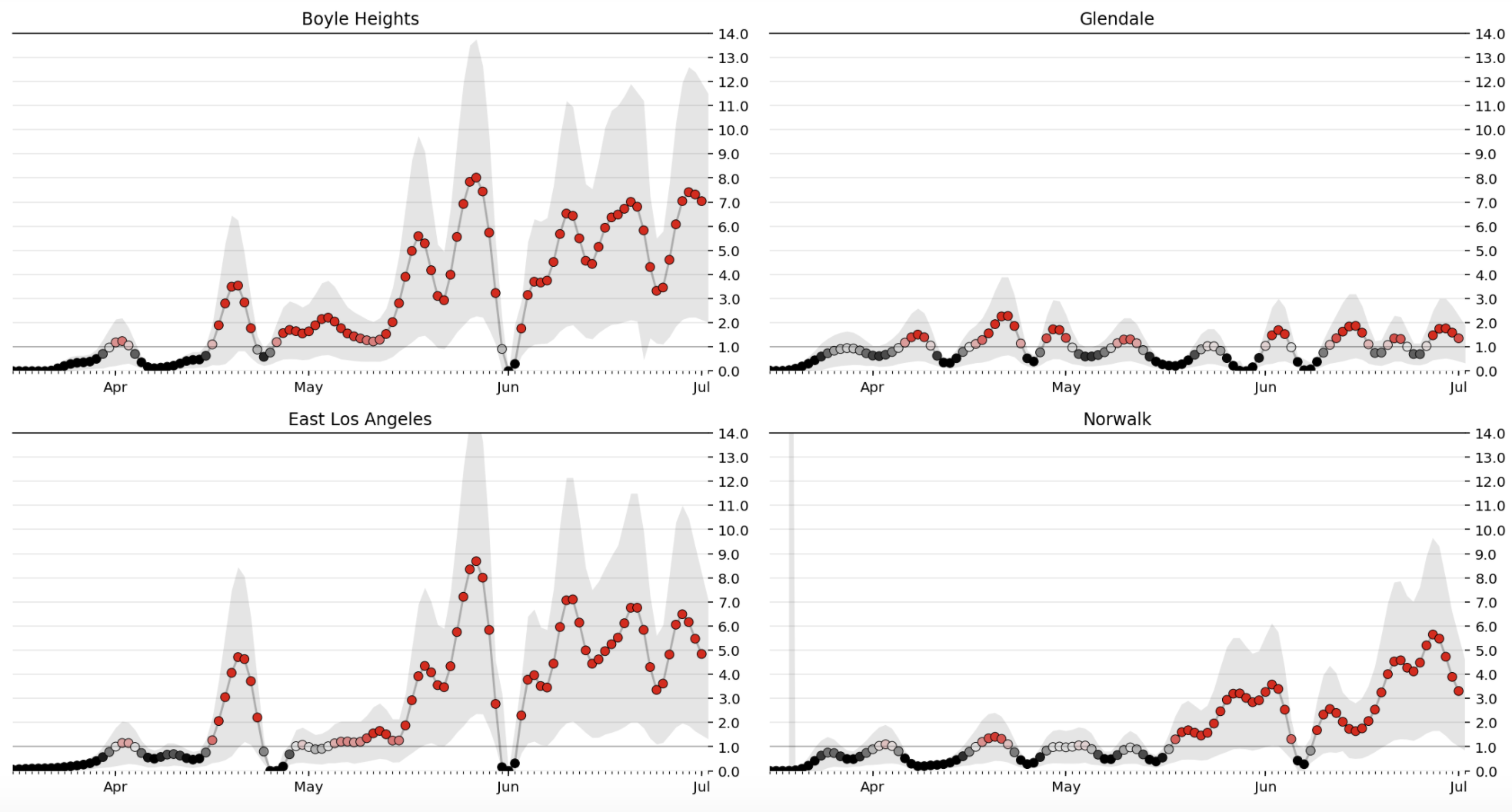}
\caption{Estimate of risk score $\Gamma_t$ over time for four representative communities in LA County: Boyle Heights, Glendale, East LA, and Norwalk. Regarding the settings, we considered the following $\mathbb{E}[D_I]=7.5$, $Var[D_I]=9$,  $\mathbb{E}[K]=3$, and $Var[K]=0.44$. Our approach also yields uncertainty in the estimate, as shown in the form of confidence intervals (in gray).}
\label{fig:rep}
\end{figure*}


\section{Implementation and Evaluation in Los Angeles County}
The software for data collection, infection rate estimation and prediction has already been implemented and made available as open-source software (at the following repository:
\url{https://github.com/ANRGUSC/covid19_risk_estimation}). The software is written in Python using standard data processing libraries such as NumPy and SciPy. 

\subsection{Data Sources}
We have acquired COVID-19 case data from the LA County's Department of Public Health using a Python-based data parser we wrote (open-sourced at the following link: \url{https://github.com/ANRGUSC/lacounty_covid19_data}). We have been updating this repository regularly with the latest data every day since mid-march and also making available plots of the number of cases, number of fatalities, top 6 communities with the large number of cases, infection rate for the entire LA County, and the top 9 communities with the highest infection rate at the following link: \url{http://anrg.usc.edu/www/covid19.html}.

The following data sources are used for the infection rate and prediction:
\begin{itemize}
\item The CoVID-19 case information was collected through LA County's daily press releases (Accessible through the following website: \url{http://publichealth.lacounty.gov/media/Coronavirus/}).
\item Recovery information provided by the World Health Organization.
\item The population data from LA County Census is available online (from \url{lacounty.gov/government/geography-statistics/cities-and-communities/}).
\end{itemize}

\subsection{Real-world Usage}
The City of Los Angeles is currently using the risk model described in this work that has been developed by researchers at USC, to help assess location-based risk for COVID-19 infection. The City is working with the County and other partners to develop a tool that is publicly accessible and can be used by individuals and communities to mitigate risk of infection. The goal is to change behaviors to reduce risk of infection and promote a greater understanding of factors that increase COVID risk. A color-coded COVID-19 threat level tool that can be used by citizens has also been unveiled by the Mayor of the City of LA, online at \url{https://corona-virus.la/covid-19-threat-level}. 

\section{Evaluation Results}
\begin{figure}
    \centering
    \includegraphics[width=5cm]{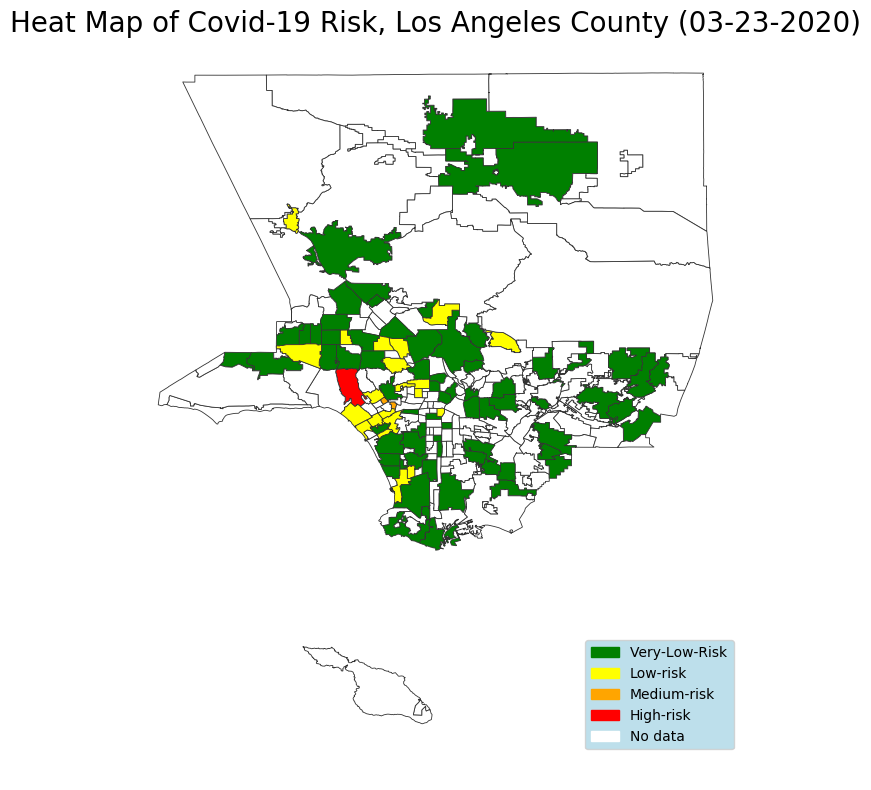} 
    \qquad
    \includegraphics[width=5cm]{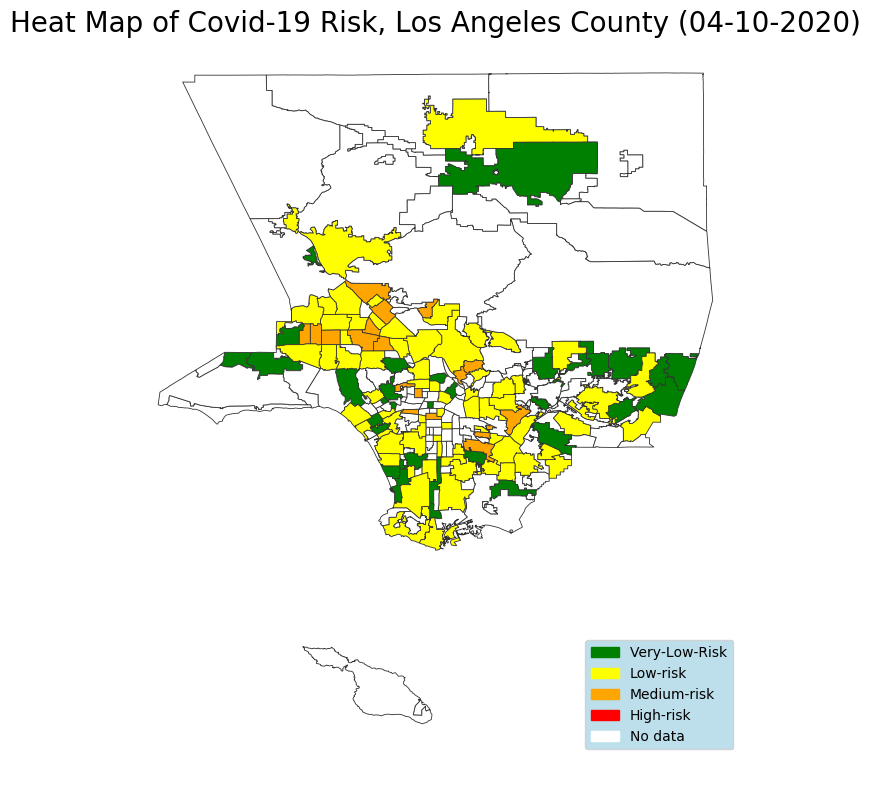}
    \includegraphics[width=5cm]{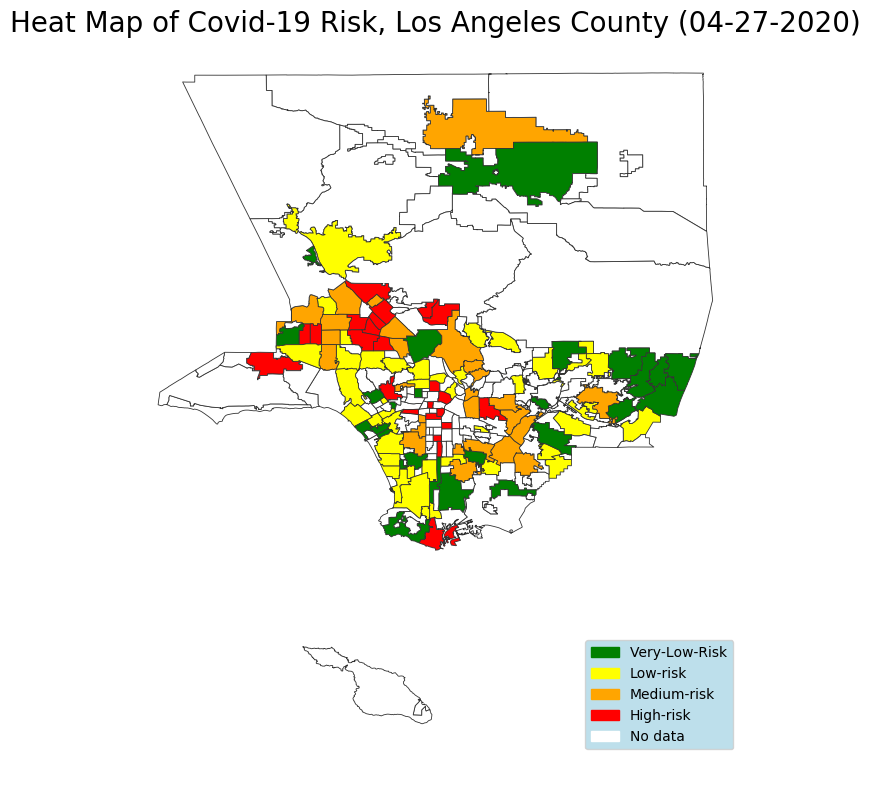}
    \newline
    \includegraphics[width=5cm]{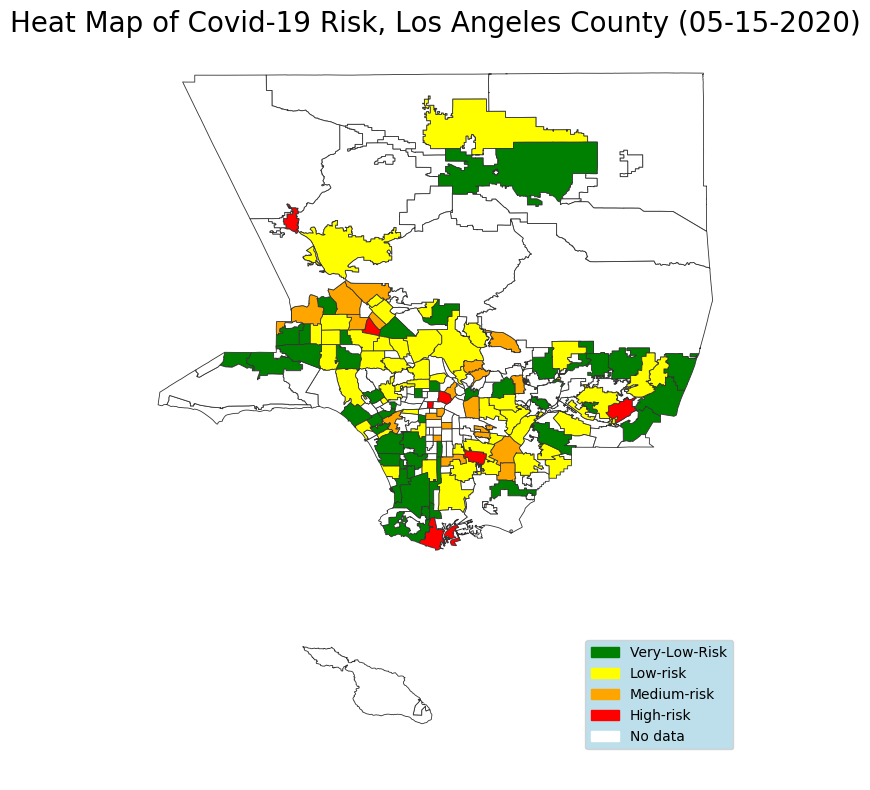}
    \qquad
    \includegraphics[width=5cm]{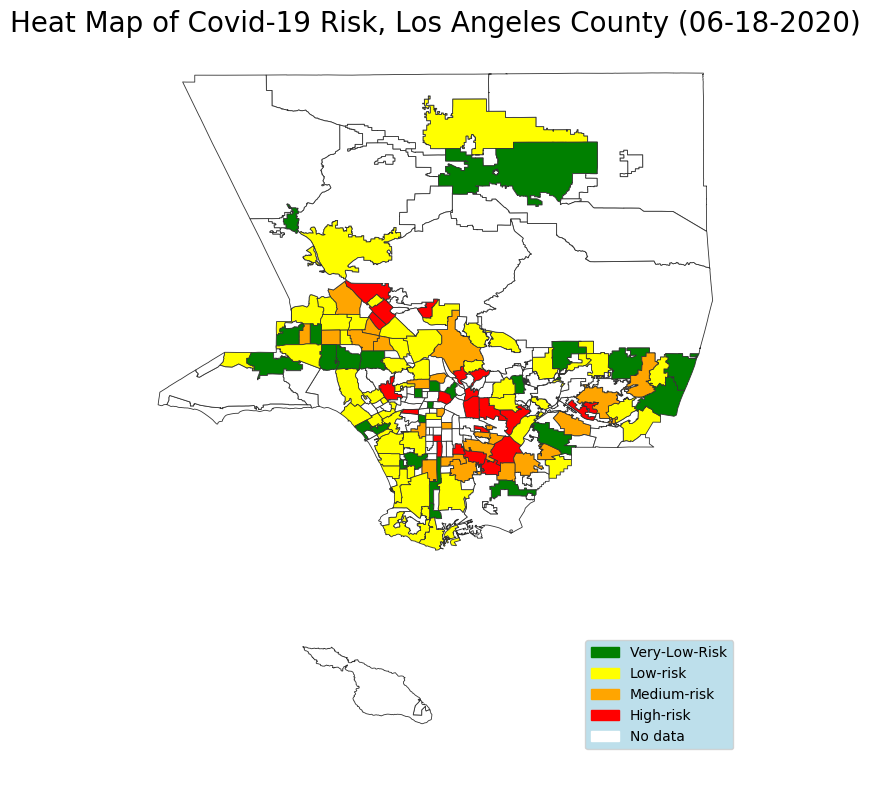}
    \includegraphics[width=5cm]{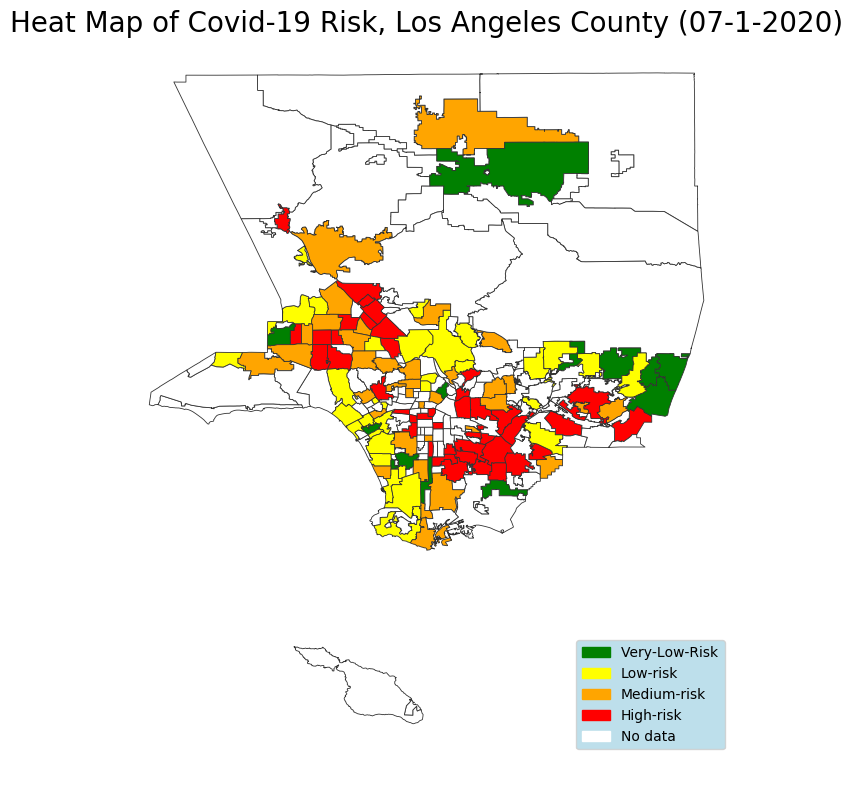}
    \caption{ Maps showing the estimated risk score for different LA County Communities on different dates since mid-March 2020. Top row: 3/23/20, 4/10/20, 4/27/20; Bottom row: 5/15/20, 6/18/20, 7/1/20}
    \label{fig:snap}
\end{figure}

We present below plots from our analysis of LA County community case data using the estimation approach described in this work. Figure~\ref{fig:entire} shows plots of the estimated expected reproductive number $R_t$ and the estimated risk score for the entire LA county. These plots are based on a 14-day moving average applied on the daily number of confirmed cases. In accordance with LA county daily press releases, there is a sharp jump in both $R_t$ and risk score around the beginning of July. Note that the reason the risk score during the beginning of July is higher than the risk score during the last week of March, despite having the same $R_t$, is due to the fact that there are significantly more confirmed cases in July compared to March. Figure~\ref{fig:rep} shows the risk score estimates over time for four representative communities within the LA County. Figure~\ref{fig:snap} shows the color-coded risk levels for communities in LA County for select dates over the past 3 months.  

\section{Conclusion}
We have proposed a new risk metric $\Gamma_t$ that can be used by individuals in any community to assess their probability of getting infected by COVID-19. The metric builds on the estimation of $R_t$, the average reproductive number, which is obtained from a time-varying extension of the classical SIR model. We show how to evaluate the uncertainty in both metrics as well. In future work, we plan to generalize the approach to the SEIR model, which also models an additional incubation period. We have released code to implement an estimation of the risk score that can be used for any community worldwide as long as time-series data for confirmed new cases and the population are known. We have also proposed the use of simple color-coded risk levels to inform and guide the public, as has been adopted in the City of Los Angeles.  

%





\ifCLASSOPTIONcaptionsoff
  \newpage
\fi



%
\bibliographystyle{IEEEtran}
\bibliography{references}




%








\end{document}